\documentclass[12pt]{article}
\usepackage{url}
\input{epsf}

\font\bm=cmmib10 at 10pt
\font\bms=cmmib10 at 7pt \textfont9=\bm \scriptfont9=\bms
\usepackage{amsfonts}
\usepackage{multirow}
\mathchardef\balpha= "790B
\mathchardef\bbeta= "790C
\mathchardef\bTheta= "7902
\mathchardef\bzeta= "7910
\mathchardef\bOmega= "790A
\mathchardef\bGamma= "7900
\mathchardef\bDelta= "7901
\mathchardef\bPhi= "7908
\mathchardef\bphi= "791E
\mathchardef\bomega= "7921
\mathchardef\bxi= "7918
\mathchardef\bet= "7911
\mathchardef\brho= "791A
\mathchardef\btau= "791C
\mathchardef\bmu= "7916
\mathchardef\bvarpi= "7924

\def \lvec{(\kern-.26em(}
\def\pmb#1{\setbox0=\hbox{#1}%
\def \lvec{(\kern-.26em(}
\kern-.025em\copy0\kern-\wd0
\kern.05em\copy0\kern-\wd0
\kern-.025em\raise.0433em\box0 }
\mathchardef\btheta= "7912

\def\XXint#1#2#3{{\setbox0=\hbox{$#1{#2#3}{\int}$}
     \vcenter{\hbox{$#2#3$}}\kern-.5\wd0}}

\usepackage{amsmath}
\usepackage{authblk}
\usepackage{url}

\usepackage{graphicx}
\usepackage{dcolumn}
\begin{document}

\title{Livestock, Methane and Climate}
\author[1]{ D. Alexander}
\author[2]{ J. D. Ferguson}
\author[3] {A. Glatzle}
\author[4] {W. Happer}
\author[5]{ W. A. van Wijngaarden}
\affil[1]{Methane Science Accord, Clevedon, New Zealand}
\affil[2]{University of Pennsylvania School of Veterinary Medicine, USA}
\affil[3] {Asociaci\'on Rural del Paraguay, Asunci\'on}
\affil[4]{Department of Physics, Princeton University, USA}
\affil[5]{Department of Physics and Astronomy, York University, Canada}

\renewcommand\Affilfont{\itshape\small}
\date{\today}
\maketitle

\begin{abstract}
Methane emissions by livestock have a negligible effect on Earth's temperature.  For example, killing all of the approximately 1.6 billion cattle on Earth in the year 2025, when this paper was written, would only reduce atmospheric methane concentrations enough to change the temperature by $\Delta T = -0.04 $ C.   Killing all 1.3 billion sheep  would lead to a temperature change of $\Delta T= -0.004 $ C.  New Zealand's pledge to reduce methane emissions of their livestock by 14\% to 24\% from those in the year 2017 would change the temperature by $\Delta T = -0.000005\hbox{ to }\!-0.000008$ C, far too small to measure. These are maximum  temperature savings where methane emissions from domestic livestock are not replaced by other sources (such as wild ruminants and termites) during the inevitable rewilding of managed grasslands and rangelands.
\end{abstract}

\section{Introduction}
Ruminant livestock like cattle, sheep, and goats, emit methane to obtain maximum nutritional energy from their forage.  Anaerobic fermentation by rumen microorganisms converts structural carbohydrates, like cellulose and hemicellulose, into nutritional organic molecules, most notably acetic, propionic and butyric acids or {\it volatile fatty acids} (VFA).  The microorganisms themselves provide proteins and other nutrients when they are digested in subsequent parts of the alimentary tract.  

An unavoidable byproduct of rumen fermentation is methane gas, CH$_4$\,\cite{methanogens}. The growth of methanogenic archaea is powered by Gibbs free energy increments released when methane molecules are formed. One  of several important methanogenic reactions\,\cite{methanogens} is 
\begin{equation}
4\hbox{H$_2$}+\hbox{CO$_2$} \rightleftharpoons\hbox{CH$_4$}+2\hbox{H$_2$O}.
\label{hm6}
\end{equation}
Under typical physiological conditions of the rumen, each CH$_4$ molecule released in reactions like (\ref{hm6}) can generate one or two energy-carrier molecules, adenosine triphosphate  (ATP).

On a per molecule basis, methane causes about 30 times more {\it radiative forcing} than carbon dioxide\,\cite{vW2020}. Nevertheless, ruminant methane emissions cause global temperature changes that are too small to measure. In this brief note we review the quantitative bases for these statements.  We will pay special attention to methane emissions by New Zealand cattle and sheep.
\section{Atmospheric Properties}
The mean surface pressure of Earth's atmosphere,  which is slightly smaller than mean sea-level pressure because of higher-altitude land surfaces, is\, \cite{pressure}
\begin{equation}
p_0=g\frac{dM}{dA}=0.985 \times 10^{5} \hbox{ N m}^{-2},
\label{int2}
\end{equation}
We assume an altitude-independent acceleration of gravity
\begin{equation}
g=9.81 \hbox{ m s}^{-2}.
\label{int4}
\end{equation}
Substituting (\ref{int4}) into (\ref{int2}) we find that the atmospheric mass per unit area is
\begin{equation}
\frac{dM}{dA}=1.004 \times 10^{4} \hbox{ kg m}^{-2}.
\label{int6}
\end{equation}
Most of the surface mass density (\ref{int6}) is due to nitrogen, oxygen and argon, but it includes a small and variable  contribution from water vapor, carbon dioxide, methane and other trace gases. We can approximate the average mass of an atmospheric molecule as\,\cite{mass}
\begin{equation}
\bar m=4.809 \times 10^{-26} \hbox{ kg}\quad\hbox{or}\quad\bar M = N_A\bar m =  28.96\hbox{ g mol}^{-1}.
\label{int8}
\end{equation}
Here one mole (mol) is Avogadro's number $N_{\rm A}$ of molecules,
\begin{equation}
N_{\rm A}=6.022\times 10^{23}.
\label{int9}
\end{equation}
The average number density of air molecules above Earth's surface is therefore
\begin{equation}
\hat N = \frac{1}{\bar m}\frac{dM}{dA}=2.088 \times 10^{29} \hbox{ m}^{-2}.
\label{int10}
\end{equation}
The mean radius of the Earth is \,\cite{Earth}
\begin{equation}
r=  6.371\times 10^{6} \hbox{ m}.
\label{int12}
\end{equation}
The total number of molecules in Earth's atmosphere is therefore
\begin{equation}
N=4\pi r^2 \hat N =1.065 \times 10^{44}.
\label{int14}
\end{equation}
\section{Methane Emissions by Cattle}
We will use the symbol $B$, from the scientific name {\it Bos taurus} or {\it Bos indicus} of the most abundant domesticated species of cattle, to denote the number of cattle on Earth.
According to reference\,\cite{global-cattle}, the total number in the year 2023 was
\begin{equation}
B(2023) = 1.576 \times 10^{9},
\label{m2}
\end{equation}
an increase from 
\begin{equation}
B(1961) =0.942 \times 10^{9},
\label{m4}
\end{equation}
in the year 1961.  
The increase has been nearly linear with time, so we will approximate the total number of cattle with the function of the calendar year $t$,
\begin{equation}
B(t) = a_B+b_Bt,
\label{m6}
\end{equation}
Eqs. (\ref{m2}) and (\ref{m4}) imply that the coefficients $a_B$ and $b_B$ of (\ref{m6}) are
\begin{equation}
a_B =-1.911\times 10^{10}\quad\hbox{and}\quad
b_B =1.023\times 10^{7}\hbox{ y}^{-1}.
\label{m8}
\end{equation}

Using the values of $a_B$ and $b_B$ of (\ref{m8}) in (\ref{m6}) we estimate the number of cattle in the year 2025 to be
\begin{equation}
B(2025) =1.6 \times 10^{9},
\label{m11}
\end{equation}
about 1 cow for every 5 of the 8.2 billion people on Earth\,\cite{worldpopulation}.

Methane emissions per cow depend on many factors: the feed intake, the feed composition, the weight and age of the cow, whether the cow is raised for meat or milk, {\it etc.}
 We will take a representative emission rate (source)  per  cow  to be\,\cite{cowmethane}
\begin{equation}
S_{B}= 70 \hbox{ kg y}^{-1}.
\label{m12}
\end{equation}
In units of molecules per unit time, (\ref{m12}) becomes 
\begin{eqnarray}
s_{B}&=&\frac{S_{B}}{m_{\rm CH_4}}\nonumber\\
&=&2.628\times 10^{27}\hbox{ y}^{-1}.
\label{m14}
\end{eqnarray}
The numerical value on the second line of (\ref{m14}) comes from the average mass of a methane molecule, CH$_4$ 
\begin{equation}
m_{\rm CH_4}=2.664\times 10^{-26} \hbox{ kg }\quad\hbox{or}\quad M_{\rm CH_4}=N_{\rm A}m_{\rm CH_4}=16.04\hbox{ g mol}^{-1}.
\label{m16}
\end{equation}

According to reference\,\cite{WikiCH4}:
\begin{quote}
The average time that a physical methane molecule is in the atmosphere is estimated to be around 9.6 years. However, the average time that the atmosphere will be affected by the emission of that molecule before reaching equilibrium – known as its `perturbation lifetime' – is approximately twelve years.
\end{quote}
We will therefore take the mean lifetime of a methane molecule in Earth's atmosphere to be
\begin{equation}
\tau=9.6  \hbox{ y}.
\label{m18}
\end{equation}
Oxidation by hydroxyl radicals, OH, to carbon dioxide, CO$_2$, and water vapor, H$_2$O, is the main destruction mechanism of methane molecules CH$_4$\,\cite{WikiCH4}. 

\begin{figure}
\begin{centering}
\includegraphics[height=90mm,width=.8\columnwidth]{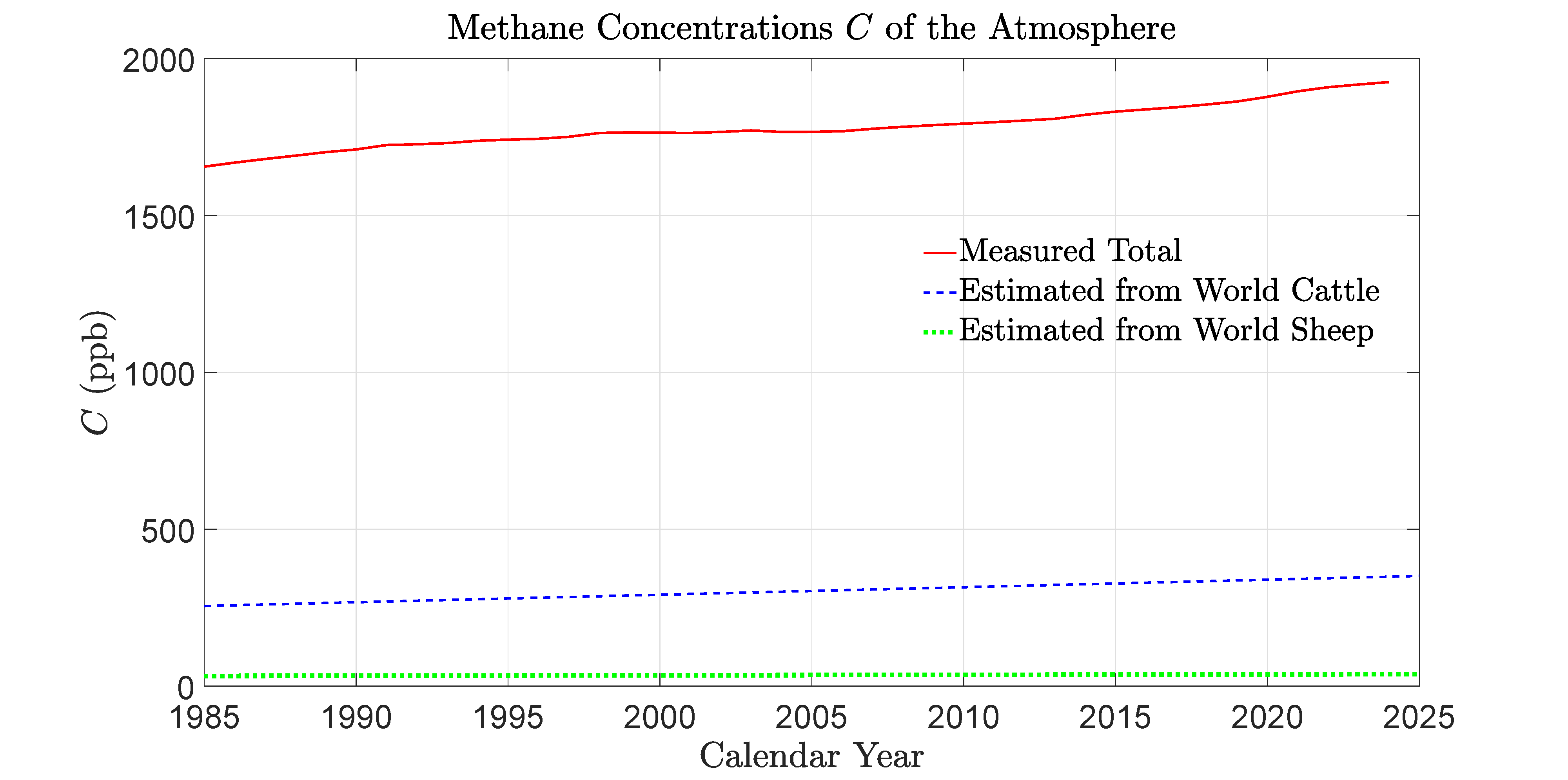}
\caption {Comparison of measured methane concentrations $C$ of the atmosphere\,\cite{WikiCH4} to the calculated fraction (\ref{m24}) from cattle and (\ref{sh26}) from sheep for the years 1985 to 2025. }
\label{CH42}
\end{centering}
\end{figure}

The number of methane molecules $N_B$ in Earth's atmosphere due to emissions from cattle will change at the rate
\begin{equation}
\frac{dN_B}{dt}=s_{B}B-\frac{1}{\tau}N_B.
\label{m20}
\end{equation}
The solution to the simple differential equation (\ref{m20}) is,
\begin{eqnarray}
N_B(t)&=&s_{B}\int_{-\infty}^t dt'B(t')e^{-(t-t')/\tau}\nonumber\\
&=&s_{B}\int_{-\infty}^t dt'[a_B+b_Bt-b_B(t-t')]e^{-(t-t')/\tau}\nonumber\\
&=&s_{B}\left[(a_B+b_B t)\tau -b_B\tau^2\right]\nonumber\\
&=&s_{B}\tau B(t-\tau).
\label{m22}
\end{eqnarray}
The first line of (\ref{m22}) gives the  number of methane molecules in the atmosphere at time $t$ due to the  cattle numbers $B(t')$ at all past times $t'$. In the subsequent lines of (\ref{m22}) we have assumed the linear dependence  (\ref{m6}) of $B(t')$ on emission time $t'$.
The dashed blue line of Fig. \ref{CH42} shows the estimated time dependence of the atmospheric methane concentration due to cattle,
\begin{eqnarray}
 C_B&=&\frac{N_B}{N}\nonumber\\
&=&\frac{s_{B}\tau B(t-\tau)}{N}.
\label{m24}
\end{eqnarray}
To evaluate (\ref{m24}) for plotting Fig. \ref{CH42}, we used the value of $s_{B}$ from (\ref{m14}), the value of $\tau$ from (\ref{m18}), the value of $N$ from (\ref{int14}), and the function $B=B(t)$ from (\ref{m6}).
Also shown in Fig. \ref{CH42} is the measured time dependence of the total methane concentration\,\cite{WikiCH4}. For the estimated cattle numbers $B$ of (\ref{m6}) and the annual methane emission per animal (\ref{m12}), the atmospheric concentration of methane due to cattle in the year $t=2025$ is
\begin{equation}
 C_B(2025)=357\hbox{ ppb},
\label{m25}
\end{equation}
 about 17\% of the atmospheric inventory.

Differentiating (\ref{m24}) and using (\ref{m6}), we find that
the rate of change of the methane concentration (\ref{m24}) due to cattle is
\begin{eqnarray}
\frac{dC_B}{dt}
&=&\left(\frac{s_{B}\tau }{N} \right)\frac{dB(t-\tau)}{dt}\nonumber\\
&=&\frac{\tau\, s_{B}\,b_B  }{N}\nonumber\\
&=&2.42 \hbox{ ppb y}^{-1}.
\label{m26}
\end{eqnarray}
The number of the last line of (\ref{m26}) came from evaluating the previous line with  (\ref{int14}), (\ref{m8}),  (\ref{m14}) and (\ref{m18}).  The relative rate of increase, from (\ref{m25})  and (\ref{m26}), is less than 1 \% per year
\begin{eqnarray}
\left(\frac{1}{C_B}\right)\frac{dC_B}{dt}= 0.68 \hbox{ \%  y}^{-1}.
\label{m28}
\end{eqnarray}
\section {Methane Emissions by Sheep}
After cattle, sheep are the next most important ruminant animal. 
We will use the symbol $O$, from the scientific name of sheep, {\it Ovis aries}, to denote the number of sheep on Earth.
According to reference\,\cite{SheepTime}, the total number of sheep in the year 2025 was
about 1.3 billion, 
\begin{eqnarray}
O(2025)&=&1.3\times 10^9,
\label{sh2}
\end{eqnarray}
an increase from about
\begin{eqnarray}
O(1961)&=&1.0\times 10^9.
\label{sh6}
\end{eqnarray}
 in the year 1961. The year-by-year fluctuations of the world sheep population, $O(t)$, have been larger than those for cattle\,\cite{global-cattle}. But a linear approximation that gives the sheep populations (\ref{sh2}) and (\ref{sh6}), and which is good enough for our purposes, is
\begin{equation}
O(t) = a_O+b_Ot,
\label{sh8}
\end{equation}
where
\begin{equation}
a_O=-8.19\times 10^{9}\quad\hbox{and}\quad
b_O =4.69\times 10^{6}\hbox{ y}^{-1}.
\label{sh10}
\end{equation}
We will take a representative emission rate per  sheep  to be 22.2 g day$^{-1}$\,\cite{sheepmethane} or approximately
\begin{equation}
S_{O}= 8.1 \hbox{ kg y}^{-1}.
\label{sh12}
\end{equation}
In units of molecules per unit time, (\ref{sh12}) becomes 
\begin{eqnarray}
s_{O}&=&\frac{S_{O}}{m_{\rm CH_4}}\nonumber\\
&=&3.06\times 10^{26}\hbox{ y}^{-1},
\label{sh14}
\end{eqnarray}
where the mass $m_{\rm CH_4}$ of a methane molecule was given by (\ref{m16}).
The dotted green line of Fig. \ref{CH42} shows the estimated atmospheric methane concentration due to sheep,
\begin{eqnarray}
 C_O&=&\frac{N_O}{N}\nonumber\\
&=&\frac{s_{O}\tau O(t-\tau)}{N}.
\label{sh26}
\end{eqnarray}

To evaluate (\ref{sh26}) for plotting Fig. \ref{CH42}, we used the value of $s_{O}$ from (\ref{sh14}), the value of $\tau$ from (\ref{m18}), the value of $N$ from (\ref{int14}), and the function $O=O(t)$ from (\ref{sh8}).
 For the estimated sheep numbers $O$ of (\ref{sh8}) and the annual methane emission per animal (\ref{sh12}), the atmospheric concentration of methane due to sheep in the year $t=2025$ is
\begin{equation}
 C_O(2025)=38\hbox{ ppb},
\label{sh28}
\end{equation}
 about 2\% of the atmospheric inventory.

Differentiating (\ref{sh26}) and using (\ref{m6}), we find that
the rate of change of the methane concentration (\ref{m24}) due to sheep is
\begin{eqnarray}
\frac{dC_O}{dt}
&=&\left(\frac{s_{O}\tau }{N} \right)\frac{dO(t-\tau)}{dt}\nonumber\\
&=&\frac{\tau\, s_{O}\,b_O }{N}\nonumber\\
&=&0.13 \hbox{ ppb y}^{-1}.
\label{sh30}
\end{eqnarray}
The number of the last line of (\ref{sh30}) came from evaluating the previous line with  (\ref{int14}), (\ref{m18}),  (\ref{sh10}) and (\ref{sh14}). The relative rate of increase, from (\ref{sh28}) and (\ref{sh30}),  is even smaller than the rate (\ref{m28}) for cattle, 
\begin{eqnarray}
\left(\frac{1}{C_O}\right)\frac{dC_O}{dt}= 0.34 \hbox{ \%  y}^{-1}.
\label{sh32}
\end{eqnarray}
\section{Radiative Forcing and Warming}
According to Eq. (36) of reference\,\cite{N2O} a small change $\Delta C$ of the atmospheric concentration of methane will cause a corresponding change $\Delta F$ of radiative forcing at the top of the atmosphere,
\begin{equation}
\Delta F=P\hat N \Delta C.
\label{rf2}
\end{equation}
In (\ref{rf2}) the column density $\hat N$ of atmospheric molecules was given by (\ref{int10}).
According to Table 3 of reference\,\cite{N2O}  the
 forcing power $P$ per added CH$_4$ molecule for the current atmosphere is
\begin{equation}
P=2.8\times 10^{-24} \hbox{ W}.
\label{rf4}
\end{equation}
According to Eq. (42) of reference\,\cite{N2O}, the forcing change, $\Delta F$, of (\ref{rf2}) would lead to a temperature change, $\Delta T$, given by
\begin{equation}
\Delta T=\Delta F\left(\frac{\partial Z}{\partial T}\right)^{-1}
\label{rf6}
\end{equation}
We will call the factor $\partial Z/\partial T$ of (\ref{rf6}) the {\it cooling capacity} of the Earth. It is the small increase,  $\partial Z$, of flux to space produced by a small increase, $\partial T$, of the average surface temperature.   As one can  see from (\ref{rf6}), for a given forcing, $\Delta F$, the temperature increase $\Delta T$ needed to maintain the cooling flux of thermal infrared radiation to space equal to the solar heating flux is inversely proportional to the cooling capacity.  
 We will use the numerical value  of the cooling capacity given by Eq. (43) of reference\,\cite{N2O}.
\begin{equation}
\left(\frac{\partial Z}{\partial T}\right)=5.6\hbox{ W m$^{-2}$ K$^{-1}$}.
\label{rf8}
\end{equation}

Since cattle  make a measurable contribution to the atmospheric methane inventory, as shown in Fig. \ref{CH42}, one might ask how much cooling would result if all cattle were destroyed, and their contribution to the atmospheric methane inventory dropped to zero.
 According to (\ref{rf2}), removing all of this methane would give a radiative forcing of
\begin{eqnarray}
\Delta F_B&=& -P \hat N  C_B(2025)\nonumber\\
&=&-0.209 \hbox{ W m$^{-2}$ }.
\label{rf12}
\end{eqnarray}
To get the number on the second line from the first, we used the value of $P$ from (\ref{rf4}), the value of $\hat N$ from (\ref{int10}) and the value of $C_B(2025)$ from (\ref{m25}).
According to (\ref{rf6}) the forcing (\ref{rf12}), due to eliminating all cattle on Earth, would cause a temperature drop of
\begin{eqnarray}
\Delta T_B&=&\Delta F_B \left(\frac{\partial Z}{\partial T}\right)^{-1}\nonumber\\
&=&-0.0373 \hbox{ C}\approx-0.04\hbox{ C}.
\label{rf14}
\end{eqnarray}
To complete the temperature drop (\ref{rf14}) would require several multiples of the atmospheric residence time, $\tau = 9.6$ y,  from (\ref{m18}).

In the year 2025, the atmospheric concentration of methane due to sheep was given by (\ref{sh28}). According to (\ref{rf2}), removing all of this methane would give a radiative forcing of
\begin{eqnarray}
\Delta F_O&=& -P \hat N  C_O(2025)\nonumber\\
&=&-0.022 \hbox{ W m$^{-2}$ }.
\label{rf18}
\end{eqnarray}
According to (\ref{rf6}) the forcing (\ref{rf18}), due to eliminating all sheep on Earth, would cause a temperature drop of
\begin{eqnarray}
\Delta T_O&=&\Delta F_O \left(\frac{\partial Z}{\partial T}\right)^{-1}\nonumber\\
&=&-0.0039\hbox{ C}\approx -0.004\hbox{ C}.
\label{rf20}
\end{eqnarray}
\section{New Zealand's Pledge}
New Zealand has pledged\,\cite{NZPledge} that by the year 2050 it will reduce methane emissions of their livestock by 14\% to 24\% below those of the year 2017.
From reference\,\cite{NZlivestock} , we estimate the number of New Zealand cattle for the year 2017 as 
\begin{equation}
B_{\rm nz}(2017) =10.1 \times 10^{6}.
\label{nz2}
\end{equation}
The number of New Zealand sheep in the year 2017 was\,\cite{NZlivestock}
\begin{equation}
O_{\rm nz}(2017) =27.5 \times 10^6.
\label{nz4}
\end{equation}
Therefore the total New Zealand emissions of methane in the year 2017 were very nearly
\begin{eqnarray}
M_{\rm nz}(2017)&=&[S_B B_{\rm nz}(2017)+S_O O_{\rm nz}(2017)]\times 1 \hbox{ y}\nonumber\\
&=&9.30\times 10^{8}\hbox { kg}
\label{nz6}
\end{eqnarray}
To get the numerical value of the second line, we used (\ref{nz2}) and (\ref{nz4}) with the numerical values of annual emissions per cow and sheep, $S_B$ and $S_O$, from (\ref{m12})  and (\ref{sh12}). Reducing these emissions by 14 - 24\% would give a concentration reduction of
\begin{eqnarray}
\Delta C _{\rm nz}&=& - [14 \hbox{ to } 24]\times 10^{-2} \frac{M_{\rm nz}(2017)}{m_{\rm CH_4} N}\nonumber\\
&=&-[4.59  \hbox{ to }  7.87] \times 10^{-11}.
\label{nz8}
\end{eqnarray}
Here the mass per methane molecule $m_{\rm CH_4}$ was given by (\ref{m16}) and the total number of atmospheric molecules $N$ was given by (\ref{int14}).
The corresponding radiative forcing (\ref{rf2}) is
\begin{eqnarray}
\Delta F_{\rm nz}&=&P\hat N \Delta C_{\rm nz} \nonumber\\
&=&-[2.68    \hbox{ to }  4.60] \times 10^{-5} \hbox{ W m}^{-2},
\label{nz10}
\end{eqnarray}
where we used (\ref{int10}) and (\ref{rf4}) to get the numbers of the second line.  The averted temperature rise follows from (\ref{rf6}) and is
\begin{eqnarray}
\Delta T _{\rm nz}&=& \left(\frac{\partial T}{\partial Z}\right)\Delta F_{\rm nz} \nonumber\\
&=&-[4.79   \hbox{ to }  8.21]\times 10^{-6} \hbox{ C},
\label{nz12}
\end{eqnarray}
less than one hundred thousandth of a degree C and much too small to measure.
\section{Summary}
Absurdly small contributions to Earth's temperature are caused by the methane emissions of ruminant livestock.  According to (\ref{rf14}), killing all the 1.6 billion cattle on Earth would cause a temperature change of about $\Delta T_B=-0.04$ C, provided that rewilding of managed grasslands and rangelands, covering around a quarter of the Earth's land surface, does not replace this source of methane emissions with other sources (such as wild ruminants and termites).  Then there would be practically no reduction of temperature.  According to (\ref{rf20})  killing all the 1.3 billion sheep on Earth would cause a temperature change of about $\Delta T_O=-0.004$ C. According to (\ref{nz4}), achieving New Zealand's goal of reducing emissions by cattle and sheep by 14\% to 24\% from those of the year 2017 would change the world temperature by about $\Delta T_{\rm nz}=-0.000005\hbox{ to } -0.000008$ C, much too small to measure.
Policies to reduce ruminant methane emissions are ``all pain, no gain."  No rational person would invest a single dollar to achieve such insignificant temperature reductions.  But some climate policies demand this.

The radiative forcing (\ref{rf4}) for methane that we use to make these estimates is slightly larger than the best IPCC values, as one can see from Table 3 of reference\, \cite{vW2020}. As discussed in reference\,\cite{N2O}, our best estimate of  the cooling capacity, $\partial Z/\partial T=5.6\hbox{ W m$^{-2}$ K$^{-1}$} $ of (\ref{rf8}), is about three times larger than those typically used by the IPCC, $\partial Z/\partial T=1.8\hbox{ W m$^{-2}$ K$^{-1}$} $, given by Eq. (48) reference\,\cite{N2O} and corresponding to strong positive feedback. But even if we use the smaller cooling capacity of the IPCC in (\ref{rf8}), the warming averted by a 14\% to 24\% reduction of methane emissions by New Zealand livestock would only be $\Delta T = -0.000015\hbox{ to }-0.000025$ C, still much too small to matter or even measure.  It is doubtful that strong positive feedbacks exist, since this would violate Le Chatelier's Principle\,\cite{LeChatelier}, that most feedbacks in natural systems are negative.


\begin{thebibliography}{99}
\bibitem{methanogens}G. Gottschalk and R. K. Thauer, {\it The Na$^{+}$-translocating methyltransferase complex from methanogenic archaea}, Biochemica et Biophysica  Acta, {\bf 1505}, 28-36 (2001).\\
\url{https://www.sciencedirect.com/science/article/pii/S0005272800002747?ref=pdf_download&fr=RR-9&rr=9af854069a58ea22}
\bibitem{vW2020} W. A. van Wijngaarden and W. Happer, {\it Dependence of Earth's Thermal Radiation on Five Most Abundant Greenhouse Gases}, Atmos. \& Oceanic Phys. arXiv: 2006.03098 (2020).
\bibitem{pressure} Mean atmospheric surface pressure,\\
\url{https://en.wikipedia.org/wiki/Atmospheric_pressure}
\bibitem{mass}Atmosphere of Earth,
\url{https://en.wikipedia.org/wiki/Atmosphere_of_Earth}
\bibitem{Earth} Earth,
\url{https://en.wikipedia.org/wiki/Earth} 
\bibitem{global-cattle}Number of Cattle,\\
\url{https://ourworldindata.org/grapher/cattle-livestock-count-heads?tab=table}
\bibitem{worldpopulation}World Population,
\url{https://worldpopulationreview.com/}
\bibitem{cowmethane}Cows and Climate Change\\
\url{https://www.ucdavis.edu/food/news/making-cattle-more-sustainable}
\bibitem{WikiCH4}Atmospheric Methane,\\
\url{https://en.wikipedia.org/wiki/Atmospheric_methane#:~:text=Atmospheric%20methane%20is%20the%20methane,the%20most%20potent%20greenhouse%20gases.}
\bibitem{SheepTime} Sheep Populations Since 1961,\\
\url{https://pulse.auctionsplus.com.au/aplus-news/insights/global-flock-versus-export-flock}
\bibitem{sheepmethane}A. Pelchen and K. J. Peters {\it Methane emissions from sheep,}
Small Ruminant Research, {\bf 26}, 137 (1998),\\
\url{https://www.sciencedirect.com/science/article/abs/pii/S092144889700031X?via%3Dihub#preview-section-abstract:~:text=A%20total%20of,on%20methane%20emissions.}
\bibitem{N2O} C. A. de Lange, J. D. Ferguson, W. Happer and W. A. van Wijngaarden,  “Nitrous Oxide and Climate”, Atmospheric and Oceanic Physics arXiv: 2211.15780   (2022).
\bibitem{NZPledge} Government resets 2050 biogenic methane targets,\\
\url{https://environment.govt.nz/news/govt-resets-2050-biogenic-methane-target/}
\bibitem{NZlivestock} Stats, NZ, {\it Agricultural production statistics: June 2017 (final)} \\
\url{https://www.stats.govt.nz/information-releases/agricultural-production-statistics-june-2017-final/}
\bibitem{LeChatelier} Le Chatelier's principle, \\ \url{https://en.wikipedia.org/wiki/Le_Chatelier%27s_principle}
\end{thebibliography}
\end{document}